\documentclass[twocolumn,showpacs,preprintnumbers,amsmath,amssymb]{revtex4}
\input epsf.sty
\usepackage{bm}
\usepackage{graphicx}
\newcommand{\bff}[1]{{\mbox{\boldmath $#1$}}}

\begin{document}
\title{Isospin dependence of nucleon effective mass in Dirac Brueckner-Hartree-Fock approach\footnote{
Correspondence author: Zhong-yu Ma\\ email:
mazy12@iris.ciae.ac.cn}}
\author{ Zhong-Yu Ma$^{a,c,d}$~, Jian Rong$^a$~, Bao-Qiu
Chen$^{a,c}$~, Zhi-Yuan Zhu$^{b,c}$ and Hong-Qiu Song$^{b,c}$ \\
\footnotesize{
$^a$ {\it China Institute of Atomic Energy, Beijing 102413, P R China}\\
$^b$ {\it Shanghai Institute of Applied Physics, Chinese
Academy of Sciences, Shanghai 201800, P R China} \\
$^{c}$ {\it Center of Theoretical Nuclear Physics, National
Laboratory of Heavy Ion Accelerator of Lanzhou, Lanzhou 730000, P
R China}\\
$^d$ {\it Institute of Theoretical Physics, Chinese Academy of
Sciences, Beijing 100080, P R China}}}
\begin{abstract}
The isospin dependence of the nucleon effective mass is
investigated in the framework of the Dirac Brueckner-Hartree-Fock
(DBHF) approach. The definition of nucleon scalar and vector
effective masses in the relativistic approach is clarified. Only
the vector effective mass is the quantity  related to the
empirical value extracted from the analysis in the nonrelatiistic
shell and optical potentials. In the relativistic mean field
theory, where the nucleon scalar and vector potentials are both
energy independent, the neutron vector potential is stronger than
that of proton in the neutron rich nuclear matter, which produces
a smaller neutron vector effective mass than that of proton. It is
pointed out that the energy dependence of nucleon potentials has
to be considered in the analysis of the isospin dependence of the
nucleon effective mass. In the DBHF the neutron vector effective
mass is larger than that of proton once the energy dependence of
nucleon potentials is considered. The results are consistent with
the analysis of phenomenological isospin dependent optical
potentials.
\end{abstract}
 \maketitle

The nucleon effective mass characterizes the propagation of a
nucleon in the nuclear medium, which is adopted to describe an
independent quasi-particle model in the nuclear many-body
system\cite{Jeu76,Mah85}. Recently, the radioactive beam physics
has become one of the frontiers in nuclear physics. It offers the
possibility to broaden our understanding of nuclear properties. In
the neutron rich side, one of the most interesting questions is
the isospin dependence of the nucleon effective interaction as
well as the nucleon effective mass. The knowledge about the
isospin dependence of the nucleon effective mass is critically
important for understanding properties of neutron stars and the
dynamics of nuclear collisions induced by radioactive
beams\cite{Riz04,Far01,LiD04}. Unfortunately, up to now the
knowledge about the isospin dependence of those quantities from
experiments is very little. Recently Li\cite{Li04} studied the
constraint of the neutron-proton effective mass splitting in the
neutron-rich nuclear matter. He found that an effective mass
splitting of $m_n^*<m_p^*$ leads to a symmetric potential that is
inconsistent with the energy dependence of the Lane potential
constrained by the nucleon-nucleus scattering experimental data.
The purpose of this work is to study the isospin dependence of the
nucleon effective mass microscopically in a relativistic approach.
It is known that there also exist some confusions on the nucleon
effective mass in the relativistic approach\cite{Jam89}. We shall
first clarify the definition of the nucleon effective mass in the
relativistic approach and, then investigate the isospin dependence
of the nucleon effective mass in the Dirac Brueckner-Hartree-Fock
(DBHF) approach.

In the non-relativistic approach the nuclear microscopic potential
$V(k,\varepsilon)$ is nonlocal and of frequency dependence. The
effective mass represents the non-locality of the underlying
microscopic nuclear potential. The non-locality of the nuclear
potential in the spatial coordinate is usually called $k$-mass
$M^*_k$ and the dynamic effect of the nuclear potential reflects
its energy dependence or non-locality in time, which corresponds
to the nucleon effective $E$-mass $M^*_E$.\cite{Jeu76,Mah85} The
effective mass is the product of those two $M^*/M=M^*_k/M \cdot
M^*_E/M $. 
The nucleon effective mass can be derived by following two
equivalent expressions\cite{Jam89},
\begin{eqnarray}\label{eq1}
  \frac{M^*}{M} &=& 1-\frac{d}{d\varepsilon}V(k(\varepsilon),\varepsilon) \nonumber \\
   &=& \left(1+\frac{M}{k}\frac{d}{dk}V(k,\varepsilon(k))\right)^{-1}_{k=k(\varepsilon)}~,
\end{eqnarray}
where $\varepsilon(k)$ is a function of the momentum $k$ defined
by the energy momentum relation,
$\varepsilon=k^2/2M+V(k,\varepsilon)$. The nucleon effective mass
can be determined from the analysis of the experimental data
performed in the framework of the nonrelativistic shell and
optical models. The typical value of the nucleon effective mass is
$M^*/M \approx 0.70 \pm 0.05$, at $\varepsilon \approx 30 $ MeV,
which is obtained by a large body of phenomenological
non-relativistic analyses of experimental scattering data.

Recently developed relativistic mean field (RMF) approach is very
successful in the description of nuclear ground state
properties\cite{Ring96}, which becomes a very popular tool in the
study of the nuclear structure and nuclear reactions. It is now
widely applied in the description of many subjects, such as exotic
nuclei, astrophysics as well as heavy nucleus collisions. In the
relativistic approach, the nucleon effective mass is always
addressed in such a way that the dressed nucleon in the nuclear
medium could be described by a Dirac equation. In the Dirac
equation the nucleon effective mass is largely reduced by the
nucleon attractive scalar potential in the nuclear medium.
\begin{equation}\label{eq2}
    M_s^*=M+U_s~,
\end{equation}
where $U_s$ is the nuclear scalar potential, which is attractive.
$M_s^*$, $M$ are the nucleon effective and bare mass,
respectively.  The nuclear scalar potential in the RMF is about
$U_s$ = -375$\pm$ 40 MeV, which corresponds to $M_s^*/M
=0.60\pm0.04$\cite{Ring96}. However, this definition of the
effective mass is not directly related to the same empirical value
derived from the analysis of experimental data in the
non-relativistic shell and optical models, as described above.

Actually the definitions in eq.(1) and eq.(2) denote different
physical quantities. This confusion has already been pointed out
and clarified by Jaminon and Mahaux\cite{Jam89} many year ago. We
may call this nucleon effective mass defined in eq.(2) the scalar
effective mass or Dirac mass\cite{Cel86}. In order to compare a
same quantity in the non-relativistic and relativistic approaches
one derives a Schroedinger equivalent equation for the upper
component of the nucleon Dirac spinor by eliminating the lower
component. The Schroedinger equivalent central potential could be
expressed as
\begin{equation}\label{eq3}
V_{cen}=(U_s^2-U_0^2+2EU_0+2MU_s)/2M~,
\end{equation}
where $U_0$ is the nucleon vector potential\cite{Ring96} and
$E=\varepsilon + M$ is the nucleon energy in the relativistic
approach.
 The Schroedinger equivalent central
potential plays the same role as the  potential $V(\varepsilon)$
in the non-relativistic approach. Therefore, the
non-relativistic-type effective mass as defined in eq.(1) can be
derived in a similar way\cite{Jam89}. In the RMF the nucleon
scalar and vector potentials are energy independent, which yields
\begin{equation}\label{eq4}
    M_v^*/M=1-U_0/M~.
\end{equation}
This effective mass may be called the vector effective mass of the
nucleon or Lorentz mass\cite{Jam89}, which characterizes the
energy dependence of the Schroedinger equivalent central
potential. Although it does not imply that any non-relativistic
limit has been taken, this definition of the effective mass could
be compared with the empirical value extracted from the analysis
of the shell and optical models. In the RMF theory the isospin
dependence of the nucleon self-energy is obtained through the
exchange of the isovector meson $\rho$, which produces a stronger
vector potential for neutron than that for proton in the
neutron-rich nuclear matter and keeps same scalar potentials for
neutron and proton. Therefore in the RMF one obtains
$M_s^*(N)=M_s^*(P)$ and $M^*_v(N)<M^*_v(P)$ in the neutron-rich
nuclear matter. If one introduces the isovector scalar meson
$\delta$ in the RMF, which produces a more attractive scalar
potential for neutron, even the neutron scalar effective mass is
smaller than that of proton in the neutron-rich nuclear matter.
However, it is well known that the microscopic nuclear scalar and
vector potentials are of the momentum and energy dependence.
Taking account of the energy dependence of the nucleon scalar and
vector potentials, the vector effective mass of a nucleon is now
obtained as follows,
\begin{eqnarray}\label{eq5}
    M_v^*/M&=&1-U_0/M-(1+U_s/M)dU_s/d\varepsilon \nonumber \\
    &&-(1+\varepsilon/M-U_0/M)dU_0/d\varepsilon~.
\end{eqnarray}

In order to consider the energy dependence of the nucleon
self-energy in the nuclear medium we shall study microscopically
the nucleon scalar and vector potentials in an asymmetric nuclear
matter(ASNM) in the DBHF approach. Recently, a new decomposition
of the Dirac structure of nucleon self-energies in the  DBHF  was
proposed by Schiller and M\"{u}ther\cite{Sch01,Uly97}. The DBHF G
matrix was separated into a bare nucleon-nucleon (NN) interaction
$V$ and a correlation term $\triangle G$. The bare NN interaction
is described by meson exchanges, such as one boson exchange
potentials (OBEP). The coupling constants and meson masses are
determined by experimental phase shifts of the NN scattering and
the ground state properties of the deuteron, which are not changed
in the DBHF approach. A projection method is applied only on the
correlation term $\triangle G$ and it is parameterized by four
pseudo-mesons. Therefore the effective NN interaction $G$ in the
DBHF in the symmetric nuclear matter (SNM) and the ASNM can be
characterized by the exchanges of those mesons in the relativistic
Hratree-Fock approach. Ma et al\cite{Mzy02} and Rong et
al\cite{Ron03} have used this scheme to analyze properties of the
ASNM and finite nuclei as well as nuclear optical potentials.
Reasonable results are achieved.
 In this work the same scheme is adopted to
investigate the effective mass of proton and neutron in the ASNM.

In consistent with the rotational invariance of the infinite
nuclear matter, the self-energy of proton and neutron can be
written as:
\begin{equation} \label{eq6}
\Sigma^{i}(k) =\Sigma_{s}^{i}(k)-\gamma_{0}\Sigma_{0}^{i}(k)+
 {\bff \gamma}\cdot {\bf k}\Sigma_{v}^{i}(k)~,
\end{equation}
where $i$ denotes proton and neutron, $\Sigma_{s}^{i}$,
$\Sigma_{0}^{i}$ and $\Sigma_{v}^{i}$ are the scalar component,
time-like and space-like parts of vector components of the nucleon
self-energy, respectively. The nucleon self-energy in nuclear
matter in the DBHF approach can be calculated with a bare NN
potential $V$ and parameterized $\triangle G$.
\begin{eqnarray} \label{eq7}
\Sigma(k) &=&
\sum_{\alpha}\int\frac{d^{4}q}{(2\pi)^{4}}\{\Gamma_{\alpha}^{a}
\Delta_{\alpha}^{ab}(0)Tr[i\Gamma_{\alpha}^{b}{\cal G}(q)] \nonumber \\
&&-i{\Gamma_{\alpha}^{a}
\Delta_{\alpha}^{ab}(q)\Gamma_{\alpha}^{b}{\cal G}(k-q)}\}~,
\end{eqnarray}
where the first and second terms are Hartree and Fock terms,
respectively. The index $\alpha$ refers to mesons, $a$ and $b$
refer to isospin components. ${\cal G}$ is the single particle
Green's function, and  $\Delta_{\alpha}^{ab}$ are meson
propagators. The vertex $\Gamma_{\alpha}^{a}$ is expressed as :
$ig$ for scalar mesons, $-g \gamma^{\mu}$ for vector mesons and
$-\frac{f}{m}\gamma^5\gamma^{\mu}q_{\mu}$ for pseudoscalar mesons,
respectively and multiplied by an isospin operator $\tau_{\alpha}$
for isovector mesons. The nucleon propagator ${\cal G}$ can be
divided into two parts: the Feynman part ${\cal G}_{F}$ and
density-dependent part ${\cal G}_{D}$. The ${\cal G}_{F}$ produces
a divergent vacuum tadpole on the nucleon self-energy, which is
neglected in the DBHF calculation. The density-dependent part
${\cal G}_{D}$ is expressed as,
\begin{equation} \label{eq8}
{\cal G}_{D}(k) =
T_{i}(\gamma^{\mu}k_{\mu}+M^{\ast}(k))\frac{i\pi}{E^{\ast}(k)}
\delta(k_{0}-\varepsilon(k))\theta(k_{F}-|{\bf k}|)~.
\end{equation}
 $T_{i}$ is the isospin part of the propagator:
\begin{equation} \label{eq9}
T_{i} = \frac{1}{2}(1\pm\tau_{3})~,
\end{equation}
where the plus and minus index correspond to proton and neutron,
respectively. The Dirac equation of a nucleon in the nuclear
medium has the form,
\begin{equation} \label{eq10}
[{\bff \alpha}\cdot {\bf
p}+\gamma_{0}(M+U_{s}^{i})+U_{0}^{i}]\psi^i({\bf r}) =
\varepsilon^i \psi^i({\bf r})~,
\end{equation}
where $U_{s}^{i}$ and $U_{0}^{i}$ are scalar and vector potential,
respectively, which are modified by the space-like vector
potential $\Sigma^{v}_{i}$.
\begin{equation}\label{eq11}
U_{s}^{i} =
\frac{\Sigma_{s}^{i}-M\Sigma_{v}^{i}}{1+\Sigma_{v}^{i}}~, ~~~~~~~~
U_{0}^{i}=
\frac{-\Sigma_{0}^{i}+\varepsilon^{i}\Sigma^{v}_{i}}{1+\Sigma^{v}_{i}}~.
\end{equation}

\begin{figure}[hbtp]
\vglue -.50cm
\includegraphics[scale=0.8]{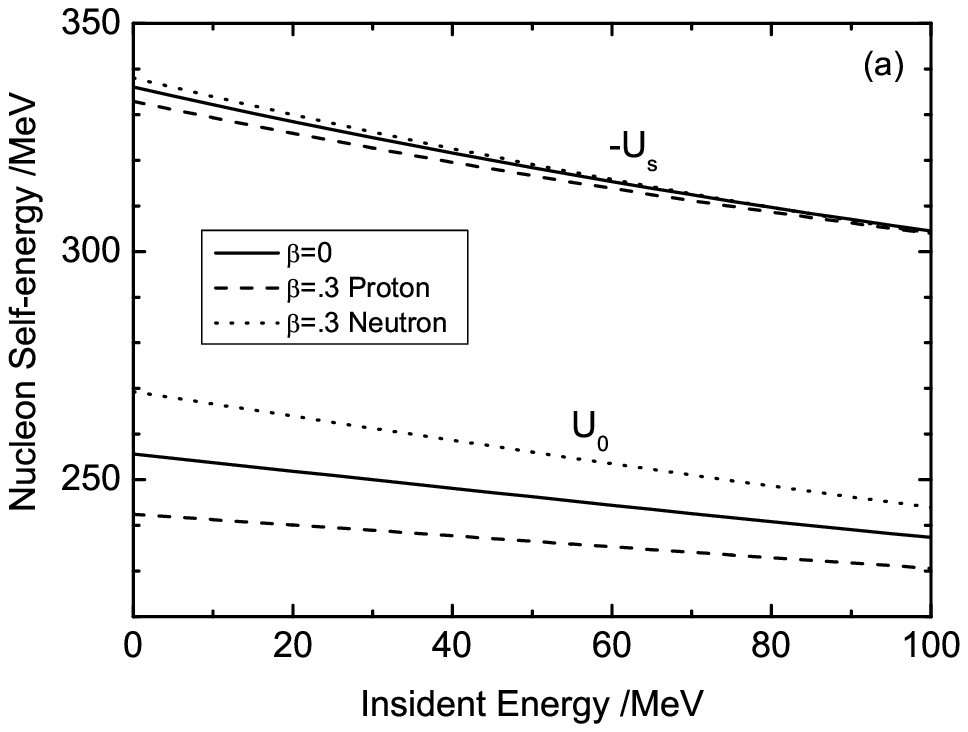}
\vglue -1cm
\includegraphics[scale=0.8]{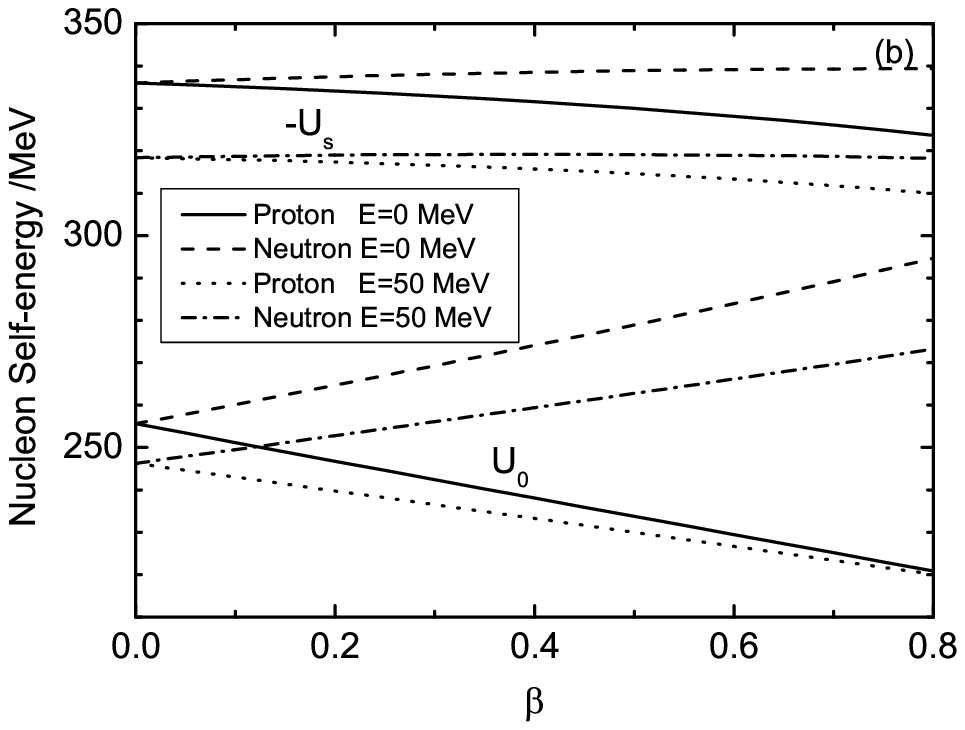}
\caption{Nucleon scalar potentials $U_s$ and vector potentials
$U_0$ of proton and neutron in the ASNM at $k_F$ = 1.36 fm$^{-1}$
as functions of the energy (a) and the asymmetry parameter (b). }
\end{figure}

In the ASNM, one defines an asymmetric parameter
\begin{equation} \label{eq9}
\beta = \frac{\rho_{n}-\rho_{p}}{\rho_{n}+\rho_{p}}~,
\end{equation}
where $\rho_{n}$ and $\rho_{p}$ denote densities of neutron and
proton, respectively. This implies that the SNM corresponds to
$\beta = 0$ and pure neutron matter to $\beta = 1.0$. The nucleon
self-energy in the ASNM  can be calculated in the DBHF. In this
work the bare NN interaction is taken as Bonn B\cite{Mac89}. The
isospin dependence of the relativistic optical potential in the
DBHF has been investigated in Ref.\cite{Ron04}, which was
consistent with the phenomenological Lane potentials extracted
from the experimental data of ($p,n$) reactions\cite{Lan62}.

The nucleon scalar potential $U_s$ and vector potential $U_0$ of
proton and neutron in the SNM and the ASNM with $\beta$=0.3 at
$k_F$ = 1.36 fm$^{-1}$ as functions of the energy are plotted in
Fig.1a. In the SNM, it is found that $U_{s}$ is about -336 MeV and
$U_{0}=256$ MeV at the Fermi surface, which correspond to the
nucleon scalar effective mass, $M^*_s/M$ = 0.64 at the Fermi
surface and the vector effective mass, $M^*_v/M$ = 0.73 if one
neglects the energy dependence of the nucleon self-energy. These
results are consistent to those obtained in Ref. \cite{Mzy02} and
cited therein. Both nucleon scalar and vector potentials decrease
as the nucleon incident energy increases. In the neutron-rich
nuclear matter the neutron self-energy, both scalar and vector
potentials become stronger than those of proton due to the larger
neutron density, while the difference of neutron and proton vector
potentials is much larger than that of scalar potentials.
Therefore the phenomenological coupling constant of the
isovector-scalar meson $\delta$ is usually taken to be rather weak
or simply omitted in the RMF. Although the energy dependence of
nucleon potentials is not strong, the energy dependence of neutron
scalar and vector potentials are steeper than those of proton,
especially for the vector potential. In the Fig.1b we plot those
potentials as functions of the asymmetry parameter $\beta$ at
nucleon incident energies $\varepsilon=$ 0 and 50 MeV. In the
neutron-rich nuclear matter neutron vector potentials become much
stronger than those for proton.

\begin{figure}[hbtp]
\vglue -.50cm
\includegraphics[scale=0.8]{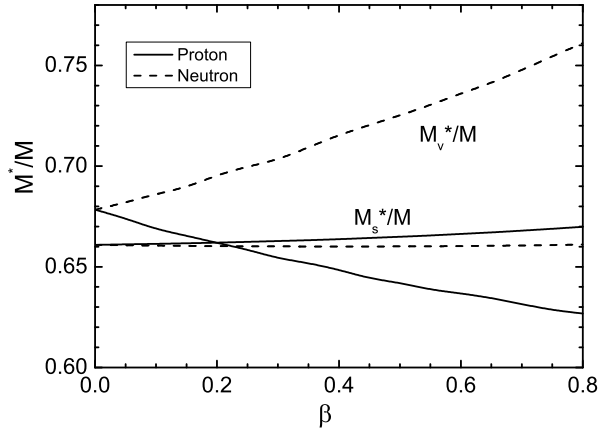}
\caption{Nucleon scalar and vector effective masses at E=50 MeV in
the asymmetric nuclear matter with $k_F$=1.36 fm.}
\end{figure}

Nucleon effective masses are calculated with those nucleon
potentials by eq.(3) and eq.(5). The results are shown in Fig.2,
where upper and lower curves correspond to nucleon scalar and
vector effective masses, respectively. The solid (dashed) curves
are those of neutron (proton). Obviously, the neutron scalar mass
is smaller than that of proton in the neutron-rich nuclear matter
due to the stronger scalar potential of neutron. However, the
difference of neutron and proton scalar effective masses is very
small in the neutron-rich nuclear matter. However, it is found
that the energy dependence of the nucleon self-energy largely
reduces the nucleon vector effective mass. At the nucleon energy
$\varepsilon$ = 50 MeV, the nucleon scalar and vector effective
masses in the SNM are 0.66 and 0.68, respectively. Due to the
stronger energy dependence of the neutron vector potential the
neutron vector effective mass becomes larger than that of proton
in the neutron-rich nuclear matter. This result is consistent with
the microscopic calculations in the non-relativistic approach,
e.g., the Landau-Fermi liquid theory\cite{Sjo76} and the BHF
approach\cite{Bom91} as well as the energy and isospin dependence
of the nucleon-nucleus scattering experimental data\cite{Li04}.

In summary, we have clarified the definition of nucleon scalar and
vector effective masses in the relativistic approach, where only
the nucleon vector effective mass could be compared with that in
the non-relativistic approach. The nucleon effective mass defined
in the non-relativistic approach could be compared with the
empirical value extracted from the analysis of the shell and
optical models. The isospin dependence of the nucleon vector
effective mass is calculated in the DBHF in the ASNM. It is found
that the neutron vector effective mass is larger than that of
proton in the neutron-rich nuclear matter if the energy dependence
of the nucleon self-energy is taken into account. This result is
consistent with what was found in the microscopic studies of
non-relativistic approach.

\begin{acknowledgments}
 This work is supported by the National Natural Science
Foundation of China under Grant Nos 10275094, 10475116, and
10235020, and the Major State Basic Research Development Programme
of China under Contract No G2000077400. MZY and CBQ would like to
thank the hospitality of the Theory group, Shanghai Institute of
Applied Physics, P.R. of China. This work was initialized during
their visit. We would also like to express special thanks to Dr.
B. A. Li for many stimulating discussions. This work has also been
supported in part by the Knowledge Innovation Project of Chinese
Academy of Sciences under grant No. KKCX2-N11.
\end{acknowledgments}

\end{document}